\documentclass{llncs}
\usepackage{makeidx}  
\usepackage[american]{babel}
\usepackage{cite}
\usepackage{url}
\usepackage{amssymb}
\usepackage{amsmath}
\usepackage{multicol,multirow}
\usepackage{color}
\usepackage{algorithm2e,algorithmicx}
\usepackage{framed}
\usepackage{graphicx}
\usepackage{caption}
\usepackage{subfigure}

\newcommand{\full}[1]{}

\begin{document}
\title{Easy 4G/LTE IMSI Catchers \\ for Non-Programmers}
\titlerunning{Easy 4G/LTE IMSI Catchers for Non-Programmers}  
%
\author{Stig F. Mj\o lsnes \and Ruxandra F.\ Olimid}
\authorrunning{} 
%
%

\institute{Department of Information Security and Communication Technology, NTNU, Norwegian University of Science and Technology, Trondheim, Norway \\
\email{sfm@ntnu.no, ruxandra.olimid@ntnu.no}
}

\maketitle              

\begin{abstract}
 
IMSI Catchers are tracking devices that break the privacy of the subscribers of mobile access networks, 
with disruptive effects to both the communication services and the trust and credibility of mobile network operators.  
Recently, we verified that IMSI Catcher attacks are really practical for the state-of-the-art 4G/LTE
mobile systems too.  Our IMSI Catcher device acquires subscription identities (\mbox{IMSIs}) within an area or location 
within a few seconds of operation and then denies access of subscribers to the commercial network.
Moreover, we demonstrate that these attack devices can be easily built and operated
using readily available tools and equipment, and without any programming. We describe our experiments
and procedures that are based on commercially available hardware and unmodified open source software.  


\end{abstract}
\section{Introduction}
\label{sec_introd}

\subsection{IMSI Catchers}
\label{subsec_imsi_catchers}

IMSI Catchers are active attack devices against the radio link protocols in mobile networks with the main goal of collecting IMSIs (International Mobile Subscriber Identities), the subscribers' identifiers used in the authentication and access control procedures.
In particular, these attacks break one of the important security requirements
set in the international specifications, namely the privacy and non-traceability of the subscriber. The malicious devices usually act as a man-in-the-middle to fulfil more advanced attacks. IMSI Catchers can be used for mass-surveillance of individuals in a geographical area, or link a real person to his/her identity in the network, trace a person with a known IMSI (check his/her presence in a building or area), eavesdrop on private conversations, all these being privacy issues concerns. IMSI disclosure is a direct consequence of the poor cryptographic mechanism used or its improper usage. 
 On the other hand, DoS (Denial-of-Service) attacks introduce serious financial losses on a targeted operator, as the subscribers cannot access the mobile services; moreover, other services that use the mobile infrastructure are impacted by the network unavailability; these include emergency calls or SMSs containing one-time codes for two-way authentication mechanisms used in the bank sector.

\subsection{Related Work}
\label{subsec_related}

IMSI Catchers were first built for 2G/GSM, and later extended to 3G/UMTS protocols. 
Until recently LTE (Long Term Evolution) protocols were considered secure against these privacy attacks due to the stronger authentication and key exchange mechanisms.  This claim has now been proved to be incorrect both by ourselves and others. Independently of our work, Shaik et al. showed that IMSI Catchers can be built for LTE  mobile networks with similar consequences as for 2G and 3G networks \cite{SBAN16}. The vulnerabilities in LTE protocols allow an adversary to trace the location of users with fine granularity, enable DoS attacks, or eavesdropping on the communication \cite{J13,SBAN16,J16,LJL16,RJP16}. General techniques to set up attacks against LTE include traffic capture, jamming and downgrade to 2G. All these academic works implement IMSI Catchers by modifying the code of open-source software projects, such as OpenLTE \cite{open_lte}, srsLTE \cite{srs_lte,G16} or gr-LTE \cite{gr_lte}. Rupprecht et al. have very recently used Open Air Interface (OAI) \cite{oai} to test compliance of UE (User Equipment) with the LTE standard with respect to encryption and to exemplify a man-in-the-middle attack, but they assume that the UE tries to connect to the rogue base station \cite{RJP16}. Most of the works use OpenLTE code because, according to these researchers, the OpenLTE architecture is easier to modify.
We use the unmodified OAI software here for the purpose of building an IMSI Catcher.

\subsection{Summary of Results} 
\label{subsec_results}

Our main goal here is to demonstrate that it is easy to build a low-cost LTE IMSI Catcher that requires absolutely no programming skills, but only readily available standard equipment and tools. Although other related works show that an IMSI Catcher for LTE can be built, they all require modification to the source code. We are the first to build an LTE IMSI Catcher using existing ``commercial off the shelf'' hardware and readily available software only. 
This result demonstrates that anyone who has basic operational computer skills (installing and running software)
can mount an IMSI Catcher attack; in particular, our attack set up requires no programming.

A second, but no less important, result is the ability of the adversary to perform a DoS attack, under the same conditions. The access to the authorised mobile network is controlled by the adversary in time and space: UEs are denied mobile services within the area of the rogue eNodeBs for the time they are active. Again, our attack set up requires no programming skills.

\section{LTE Mobile Networks}
\label{sec_lte}

\subsection{LTE Architecture}
\label{subsec_lte_infra}

\begin{figure}[t]
\centering

\includegraphics[scale=0.4]{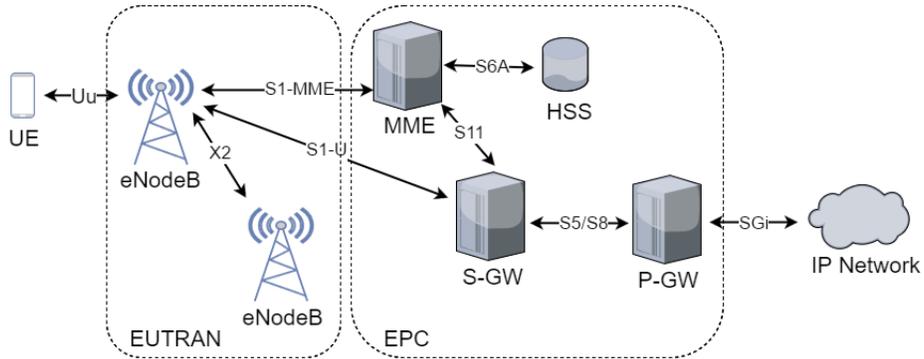}

\caption{LTE Architecture}
\label{fig_lte}

\end{figure}

LTE architecture (see Figure \ref{fig_lte}) consists of two parts: EUTRAN (Evolved Universal Terrestrial Radio Access Network) and EPC (Evolved Packet Core), each of them being composed of several components that we briefly described next. 

\textit{UE (User Equipment)} refers to the mobile user terminal. It contains a USIM (Universal Subscriber Identity Module), which stores the IMSI (International Mobile Subscriber Identity) and the associated permanent secret key, used to derive temporary keys for authentication and encryption. IMSI is a permanent unique value that globally identifies the subscriber. 

\textit{eNodeB (evolved NodeB)} is the base station that communicates with UEs by radio links and represents the access point to the operator's network.

\textit{MME (Mobile Management Entity)} is responsible for authentication and resources allocation to UEs. It manages the mobility of UEs in the network when eNodeBs cannot.

 \textit{HSS (Home Subscriber Server)} stores the authentication parameters, private keys and other details about the UEs. 
 
 \textit{S-GW (Serving Gateway)} is an interconnection point between EUTRAN and EPC, being and anchor point  for intra-LTE mobility.

 \textit{P-GW (PDN Gateway)} is a routing point to provide connectivity to the external PDN(Packet Data Network).
 
\medskip
A mobile network is identified by MCC (Mobile Country Code) and MNC (Mobile Network Code). The MCC and MNC are publicly available online \cite{mcc_mnc}.
An eNodeB controls a set of cells running at a specific frequency in an LTE band, or equivalently, on an EARFCN (EUTRA Absolute Radio-Frequency Channel Number)  \cite{lte_freq_band,lte_calculator}. The frequency ranges allocated to the network operators are also public information \cite{freq_list}.
Several cells in a geographically region form a TAC (Tracking Area Code), which is managed by a single MME.
The eNodeB broadcasts information such as MCC, MNC or TAC via SIB (System Information Block) messages. This allows the UE to identify the operator. 
To attach to the network, UE then sends an \textsf{ATTACH\_REQUEST} message to the eNodeB. If UE moves to a new area, it initiates a \textsf{TAU\_REQUEST} and a \textsf{TRACKING\_AREA\_UPDATE} procedure is performed.
 
We skip other details here, but invite the reader to refer to the specifications for more details \cite{ETSI_331,ETSI_301}.

\subsection{LTE Cell Selection and Reselection} 
\label{subsec_lte_cell}

A mobile device that attempts to access the mobile network performs a \textit{cell selection} procedure. This means that the UE searches for a suitable cell and chooses to camp on the one that provides the best service. If later on the UE finds a more suitable cell (according to some reselection criteria), then it performs \textit{cell reselection} and camps onto the new cell.
Cell selection and reselection in LTE mobile networks are complex processes involving several steps and different criteria \cite{ETSI_304,ETSI_133}. We do not describe the complete processes here, but only highlight the aspects that are further required for the understanding of our work.

When the UE is switched on, it camps on a cell within a PLMN (Public Land Mobile Network) selected accordingly to a list stored locally and some selection criteria. In practice, the selected PLMN corresponds to the last mobile network the UE had successfully connected to before switch off. Note that this is not always the home PLMN (the mobile network the USIM belongs to).

At reselection, the UE monitors intra-frequency, inter-frequency and inter-RAT (Radio Access Technology) cells indicated by the serving cell.  
In the following, we refer to inter-frequency reselection, as its understanding is a prerequisite for our results. The UE performs inter-frequency reselection when it camps on a cell that runs on a different frequency than the serving cell. First level criterion for reselection is the absolute priority list: the UE always monitors and tries to camp on cells that run on higher priority frequencies. The eNodeB of the serving cell broadcasts in clear the list of absolute priorities (along with other reselection parameters) in SIB messages. For each frequency that is listed, \textsf{cellReselectionPriority} field defines its absolute priority, where 0 indicates the lowest priority and 7 indicates the highest priority.
Note that LTE reselection uses the radio link quality as second level criteria, so simply running a rogue eNodeB in the vicinity of the UE does not automatically trigger a reselection. 

Operators use the absolute priority frequency criteria to gain robust coverage. They design the network such that within an area coexist multiple cells that run on distinct frequencies with associated priorities. In case of incidents or network congestion on the highest priority frequency, the UE will not lose connectivity, but reselect a cell that runs on the next priority frequency.
Deciding the number of frequency levels that should coexist in an area and their associated priorities is a task of the design network engineer.

\subsection{Security Requirements and the Adversarial Model}
\label{subsec_adv_model}

The main goals of the adversary are to collect IMSIs and to deny network availability.
The adversary is constrained in both budget and technical skills, so it aims for a low-cost and simple attack that only requires basic computer knowledge. Our adversary will only use commercially available hardware and unmodified open source software.

Furthermore, the adversarial model follows the one of previous work \cite{SBAN16,J16}. The attacker is active in the sense that it can build up a rogue eNodeB that interacts with the UEs. The model of a passive adversary (an adversary that can only sniff LTE traffic, but cannot actively interfere) is inappropriate both for practical and theoretical reasons. On the practical side, active adversaries have been proved successful, so LTE mobile networks must protect against them, while on the theoretical side TMSIs (Temporary Mobile Subscriber Identities) have been introduced as a security measure starting with 2G.
TMSIs are temporary values that replace the IMSIs in the process of subscribers' identification, minimising the transmission of IMSIs through the air and hence drastically decreasing the chances of a passive adversary to collect them.

\section{Equipment and Tools}
\label{sec_exp_setup}

\subsection{Hardware} 
\label{subsec_exp_setup_hard}

\begin{figure}[b!]
\centering

\includegraphics[scale=0.055]{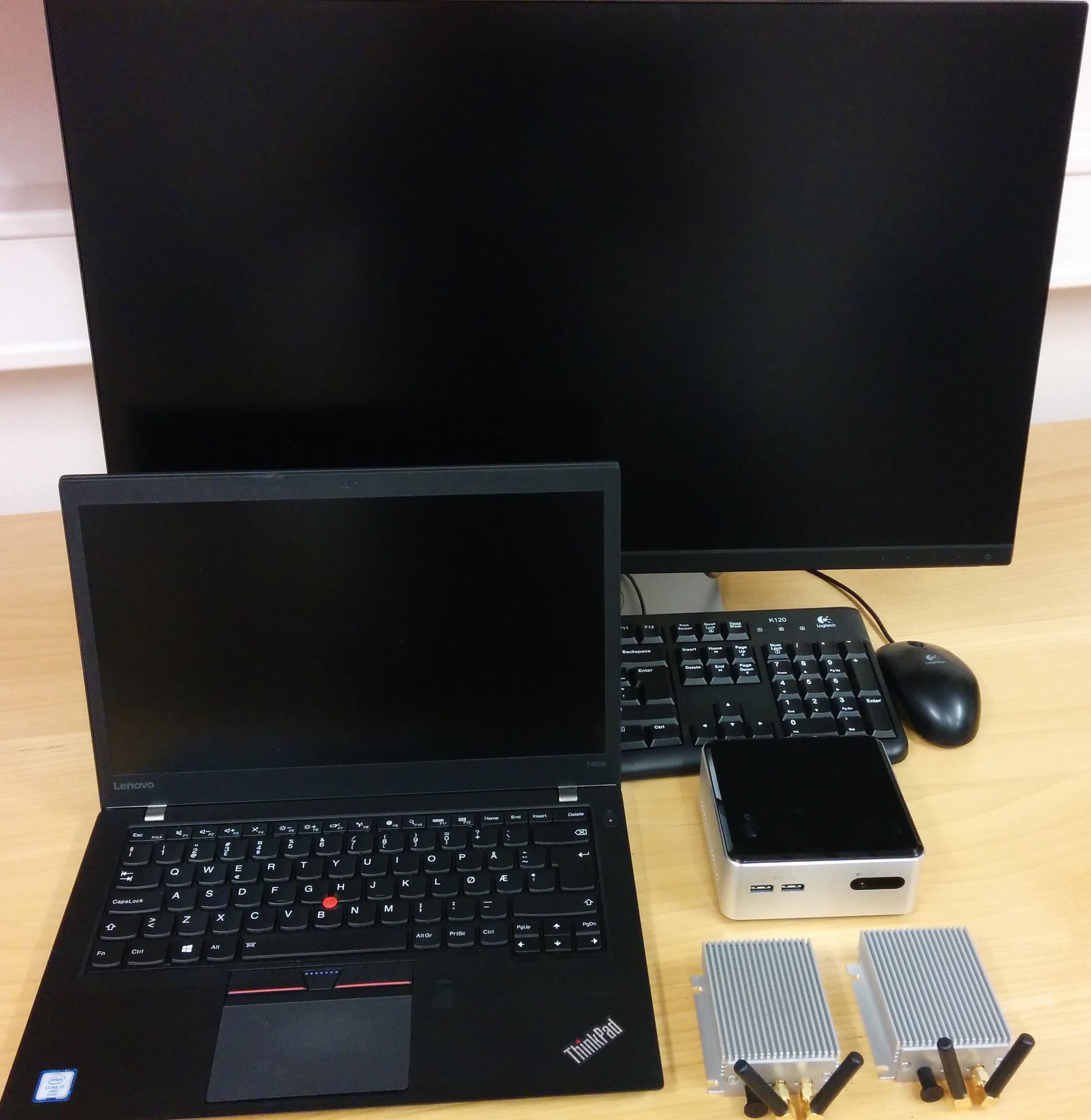}

\caption{Experimental Hardware (excluding UEs)}
\label{fig_testbed}

\end{figure}

Only Commercial Off-The-Shelf (COTS) devices have been used in our
attack setup. The hardware components that we used for our experiments
(excluding the mobile phones) are
shown in Figure \ref{fig_testbed}.

\textbf{Computers.} We used two different computers: 
one Intel NUC D54250WYK (i5-4250U CPU@1.30GHz) and one Lenovo ThinkPad T460s (i7-6600U CPU@ 2,30GHz). 
Both run 64-bit Kubuntu 14.04 kernel version 3.19.0-61-low latency and have USB3 ports, 
which are prerequisites for running the OAI software.  
The Intel NUC computer was attached with standard peripherals (display screen, mouse, keyboard).

\textbf{USRPs.} The B200mini is a USRP (Universal Software Radio
Peripheral) from Ettus Research that can be programmed to operate over a wide
radio-frequency range (70MHz - 6GHz), communicating in full duplex. It
can be used for all of the LTE frequency bands. The technical
specifications of the B200mini are available at \cite{B200mini}.
We used two USRP B200mini to set up the eNodeBs. 

\textbf{User Equipment.} We used a Samsung Galaxy S4 device 
to find the LTE channels and TACs used in the targeted area.
For testing the IMSI Catcher, we used two LG Nexus 5X phones running
Android v6.  We used different USIMs from the two biggest
Norwegian operators, and the two biggest Romanian operators
(see Section \ref{sec_results}).

\bigskip The total hardware cost is less than 3000EUR. Any two OAI
compatible computers and USRPs can be used to mount the experiments \cite{oai_hw_req}. 
Also, a single mobile device is sufficient if it can provide the minimal information
needed (EARFCN and TAC) for setting up the rogue eNodeBs. So, the
minimal kit includes: two computers and two USRPs for the network side
and one mobile device with a USIM from the targeted mobile operator for
the client side. Keeping the configuration to minimum, cost might be decreased.  Also, we expect the costs to be significantly lower in the near future.


\subsection{Software}
\label{subsec_exp_setup_soft}

We used open-source freely available software that
did not require any modification for our purposes to build and run the 4G/LTE IMSI Catcher.

\textbf{Service and Testing Modes.} 
Seen as a facility of the operating system and the privilege rights of
the user, service or testing modes of mobile devices offer important
information about the LTE network. We describe, for comparison, the information displayed by the two
types of phones we used during the experiments

\begin{figure}[b!]
\centering

\subfigure[Service Mode \newline (Samsung, root)]{\label{fig_samsung_1}\includegraphics[scale=0.21]{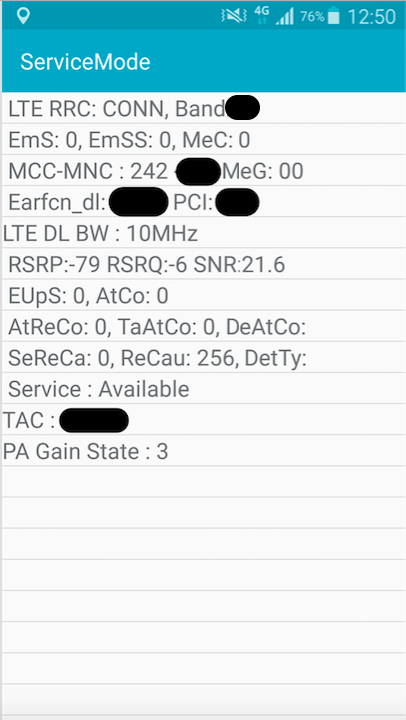}}
\hspace{.5in}
\subfigure[Testing Mode \newline (Android, non-root)]{\label{fig_android}\includegraphics[scale=0.21]{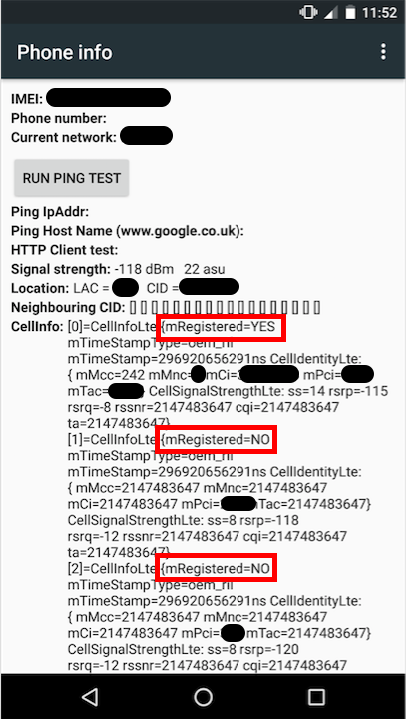}}
\caption{Service and Testing Modes}
\label{fig:test}
\end{figure}

\textit{Samsung phones} offer LTE connection details by default
\cite{samsung}. To access Service Mode, call *\#0011\#. The most
important pieces of information are the EARFCN\_DL (downlink EARFCN) and the TAC. Other interesting information include the MCC,
MNC and Cell ID. Refer to Figure \ref{fig_samsung_1} for more details.

\textit{Android phones} (including Samsung phones) access Testing Mode
by calling *\#*\#4636\#*\#*. This is a feature available on all Android
phones, which does not necessarily display EARFCN\_DL by default (it is dependent of the Android version), but displays
information about several LTE cells that coexist in the area on which the
phone might downgrade to in case the actual cell becomes unavailable.
Figure \ref{fig_android} shows the Testing Mode of a LG Nexus 5X
display. The UE is registered to the first displayed cell
(mRegistered=YES), while the others are showed to be accessible.
However, last versions of Android or applications such as LTE G-Net Track Lite or NetCell
Tracker \cite{app1,app2} (in root mode) can provide EARFCN\_DL and other information in
a user-friendly format.

\textbf{OAI (Open Air Interface).} 
OAI  is open source software that implements both the core network (EPC) and
access-network (EUTRAN) of 3GPP cellular networks \cite{oai,NKK14}.
Currently, it provides a standard compliant implementation of Release 10
LTE and establishes generic interfaces with various hardware
manufactures (including Ettus Research). Basically, the LTE network is
emulated on a computer, and the USRP is used as the radio platform for
the eNodeB implementation.
It is recommended to run EPC and EUTRAN on different machines, but OAI
accepts both on a single computer too. For our tests, keeping in mind
the goal of a low-cost IMSI Catcher, we used two machines,
one for each of the two rogue eNodeBs that must run
concurrently. 

\textbf{OpenBTS.} 
OpenBTS is an open source software development project for GSM mobile networks
\cite{open_bts}. OpenRegistration in OpenBTS allows an IMSI that match a
given regular expression to access the network. The one-sided authentication protocol of GSM
permits the OpenBTS process to accept mobile terminals without proper mutual authentication. Note that we have used OpenBTS for testing purposes only, to make a roaming USIM 
connect to a masquerade home network, but this is not a prerequisite for
setting up the LTE IMSI Catcher (see Section~\ref{subsec_imsi_catcher_roaming}).

\section{The LTE IMSI Catcher}
\label{sec_imsi_catcher}

We now describe how we built and operated our LTE IMSI Catcher. We ran
two rogue eNodeBs using the OAI software, and
set up with the proper configuration files. One rogue eNodeB (called
\textsf{eNodeB\_Jammer} from now on) causes the UE to detach from the
serving cell that it is camped on, and to reselect to our rogue cell set up by the second eNodeB (called
\textsf{eNodeB\_Collector}), which masquerades as an authorized eNodeB
but with higher signal power.  

The \textsf{eNodeB\_Collector} broadcasts the MCC and the MNC of the
target network operator to impersonate the real network. The \textsf{eNodeB\_Collector} signals
a TAC value different from the commercial one, which brings the UE to
initiate a \textsf{TAU\_REQUEST} message. For simplicity, 
we configured it to the next
available TAC (TAC of the commercial network + 1).  We found that
 the TAC must not be changed for multiple runs of the experiment (assuming the
commercial TAC is unchanged), therefore we kept this value constant. Besides the
MCC-MNC of the target network, the \textsf{eNodeB\_Collector} must run
on the LTE EARFCN (the absolute physical radio channel)
 which corresponds to the highest priority next to the
jammed channel. This assures that the UE prefers the
\textsf{eNodeB\_Collector} prior to any other commercial eNodeB in the
area. The \textsf{eNodeB\_Jammer} is turned on, using
the radio channel of the cell that the UE camps on. This jams the radio interface and
decreases the signal of the commercial eNodeB under the specified
threshold (see Section \ref{sec_lte}), causing the UE to trigger
a new search for available eNodeBs. The UE tries to camp to the cell
that runs on the next priority frequency and provides the best signal,
namely the rogue eNodeB.

We divide the adversarial actions into two main phases:

\paragraph{Phase 1. Gather the configuration parameters (EARFCN and
TAC):}
\begin{enumerate} \item Connect a mobile phone to the target
network and read the EARFCN\_DL and TAC; \item Configure and run the
\textsf{eNodeB\_Jammer}, using the MCC and MNC of the target network and
the EARFCN\_DL from the previous step; \item Read the new value of
EARFCN\_DL, after the UE performs reselection. 
\end{enumerate}

\paragraph{Phase 2. Configure and run the LTE IMSI Catcher:}
\begin{enumerate} \item Configure and run the
\textsf{eNodeB\_Collector}, using the MCC and MNC of the target network,
a different TAC than the one in \textit{Phase 1.1} and the EARFCN\_DL set to
the value in \textit{Phase 1.3}; \item Configure and run the
\textsf{eNodeB\_Jammer}, using the MCC and MNC of the target network and
the EARFCN\_DL set to the value in \textit{Phase 1.1}.
\end{enumerate}

The channel displayed in \textit{Phase 1.1} is associated with the highest
priority (unless the signal power is below the reselection
threshold, see Section \ref{subsec_lte_cell}). The UE connects to it
even if the signal power is not so strong. This can be easily seen by
comparing the information displayed by the mobile device before and
after the \textsf{eNodeB\_Jammer} is turned on. The channel in
\textit{Phase 1.3} has either the same priority, but lower signal power,
or lower priority, regardless the signal power. Once the
\textsf{eNodeB\_Jammer} is active, this triggers an
\textsf{ATTACH\_REQUEST} message from the UE to the \textsf{eNodeB\_Collector}.
Then the UE will reveal its IMSI as a response to an \textsf{IDENTITY\_REQUEST} query
from our Collector cell.

\section{Experiments and Results}
\label{sec_results}

\subsection{Introduction}
\label{subsec_introduction}

All our experiments were run in our wireless security lab at the university.
We intend our work to be a motivation for solving the problem of privacy attacks in mobile networks 
both in existing and emerging systems at the technical specification level, 
as well as for mobile network operators to use the already existing security mechanisms. 

\subsection{IMSI Catching}
\label{subsec_imsi_catcher}

We successfully ran experiments with three subscriptions from two mobile
operators in Norway, testing two, respectively one USIM card for each.
For all, we used the same mobile device, the LG Nexus 5X.
To respect privacy of other users, IMSI values were captured from our own USIM cards only.

\begin{figure}[t!]
\centering

\includegraphics[scale=0.3]{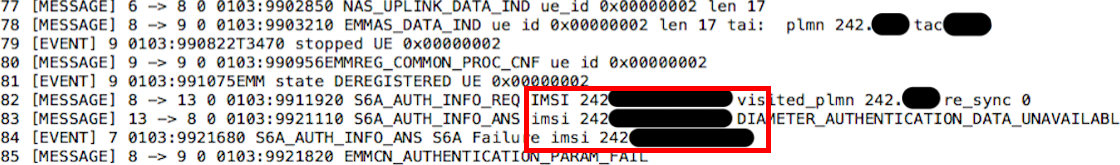}

\caption{IMSI Capture}
\label{fig_imsi_catcher}

\end{figure}

All the three
IMSIs used for tests were successfully captured by the \textsf{eNodeB\_Collector}
within a few seconds after the \textsf{eNodeB\_Jammer} is switched on.  Figure
\ref{fig_imsi_catcher} shows a portion from the log file that contains a
captured IMSI.

Experiments were repeated several times, with different values for Cell ID
or eNodeB ID. They were all successful, so we conclude there is no
protection mechanism in place (e.g. a list of accepted cells, TACs or
timers). 

\subsection{Denial-of-Service} 
\label{subsec_imsi_catcher_dos}

\begin{figure}[t!]
\centering

\includegraphics[scale=0.33]{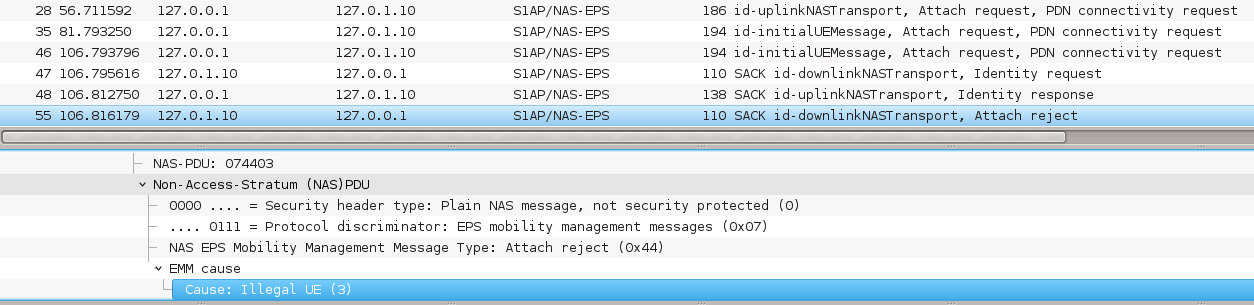}

\caption{\textsf{ATTACH\_REJECT} Message}
\label{fig_pcap}

\end{figure}

We observed that running the IMSI Catcher results in a DoS
attack because authentication fails after the UE triggers the
\textsf{ATTACH\_REQUEST} and consequently sends its identity as a
response to the \textsf{IDENTITY\_REQUEST}. This is normal, as the IMSI
is not recognised by the HSS as a valid subscriber and no handover
procedures can be accomplished. The \textsf{eNodeB\_Collector} responds
to the UE with \textsf{ATTACH\_REJECT} cause 3 (Illegal
UE), making the UE to consider the network unavailable until reboot
(Figure \ref{fig_pcap}) \cite{ETSI_301}.\footnote{Depending on the exact software version of OAI being used, UE connectivity to the eRogueB fails in various ways, but all end up with DoS until the reboot of UE.}

If the UE is powered off and on again while our rogue eNodeBs are still up,
the UE keeps failing to camp on any cell. This turns
into a controlled DoS attack against the target network in the area covered.
This DoS mode remains active as long as the rogue eNodeBs are up. 

\subsection{Roaming}
\label{subsec_imsi_catcher_roaming}

An even simpler IMSI Catcher is available for USIMs in a roaming situation, 
in particular under restricted scenarios like airport arrivals. 
The adversary can now run a single rogue eNodeB with the MCC and MNC of the
home network.

We distinguish 2 scenarios: (1) USIMs that are inactive in roaming (roaming was not activated or the user travels to a country without any roaming agreements with the home network) and (2) USIMs that are active in roaming, but did not connect to any network while in roaming. For both cases, the UE reveals its IMSI on power on, if the IMSI Catcher is up and running at that time. 

\medskip \textbf{USIMs inactive in roaming.} A USIM for which roaming is disabled
is directly susceptible to an LTE IMSI Catcher using the
same MCC and MNC as the home network. Hence, to setup an IMSI Catcher in this
scenario, an adversary simply configures the corresponding MNC and MCC,
then runs OAI. The LTE frequency band, EARFCN or any other information
are irrelevant in this scenario.  If the IMSI Catcher is running when the UE
is turned on, then the UE will try to camp on the cell that masquerades
the HPLMN and hence, responds to the \textsf{IDENTITY\_REQUEST} message
with its IMSI.

\begin{figure}[t]
\centering

\includegraphics[scale=0.37]{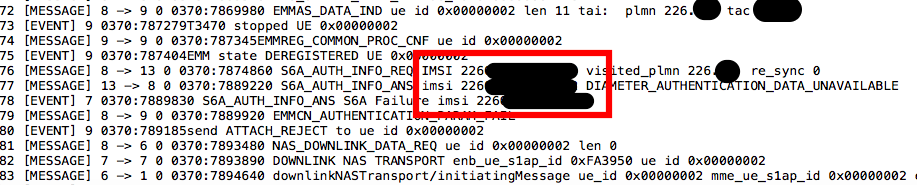}

\caption{IMSI Capture in Roaming}
\label{fig_roaming}

\end{figure}

We carried out this attack using an inactive LTE USIM card from one of
the most known Romanian mobile operators. Figure \ref{fig_roaming} shows
the IMSI, as captured by the IMSI Catcher as a response to an
\textsf{IDENTITY\_REQUEST} message.

\medskip \textbf{USIM active in roaming.}  A USIM for which roaming is enabled is
susceptible to an IMSI Catcher built as previously
explained, when the UE is first powered on in roaming. 
Once connected in
roaming, the UE will not search for its home network because in
practice, regardless of the specifications, the UE will always try to
connect to the most recent network for which the connection was successfully
(this is an issue inherited from previous generations, as for example it
exists in GPRS also \cite{M05}). The scenario could make sense on an
airport: the user turns off its device (or equivalently, enable the fly
mode) while connected to the home network and later turns it on (or
equivalently, disables the fly mode) after landing, in the presence of
a rogue eNodeB that simulates the home network. The UE will try to
connect to the rogue eNodeB and hence, it will reveal its IMSI as a
response to the \textsf{IDENTITY\_REQUEST} message. Such an IMSI Catcher
might be used by agencies that want
to survey the presence of people arriving from specific countries to
their own territory in a hidden manner (without revealing their
intentions by asking official documents from the border authorities).
Independent work considers this scenario in theory and proposes
correctness and completeness of location update trails and
geographically plausibility of location updates to address this issue 
\cite{DPW16}. We have hereby certified this functionality.

We performed the experiment by using an active LTE USIM card belonging to
another network operator in Romania. As the USIM had been already active
in roaming, we simulated the test case by connecting it to a fictive 2G
network that masquerades the home network. For this, we used OpenBTS in
OpenRegistration mode. After the UE connected to the GSM network, we
switched it to 4G and immediately turned it off. At power on, the UE tried to
connect to the rogue eNodeB that simulated the home 4G network and then
revealed its IMSI.

\section{Conclusion}
\label{sec_future_work}

In conclusion, our experiments have verified, independently of other works, that IMSI-catching indeed can be done for the 4G/LTE system too.
We claim that (1) IMSIs can be collected by the \textsf{eNodeB\_Collector}, and
(2) DoS (Denial-of-Service) is performed automatically after the UE receives an attach
rejection message (with a specific cause). The 4G/LTE is vulnerable to active privacy attacks by IMSI
Catcher, and we found that these attacks
can be done quite easily and therefore can impact the confidence and reliability of 
commercial mobile networks. We showed that these attacks
are not limited to clever programmers with special hardware. We hope that this report,
and others, will make the 4G/LTE service providers aware of this threat, and lead to demands for improved
privacy and security protocols in the mobile networks.

\subsubsection*{Acknowledgements.}

The authors would like to thank master student Fredrik Skretteberg 
for providing the Samsung phone necessary for some experiments.

\bibliographystyle{splncs}
\bibliography{llncs}

\end{document}